# Impact of high renewable penetration scenarios on system reliability: two case studies in the Kingdom of Saudi Arabia

Markus Groissböck and Alexandre Gusmao, Saudi Aramco (PSPD/PSRD), Dhahran, Saudi Arabia

email: markus.groissboeck@aramco.com / alexander.gusmao@aramco.com

ORCID: https://orcid.org/0000-0001-6025-9477 / n/a

*Abstract* — **Renewable energy sources (RES) are close to grid parity compared with conventional generation. Therefore, it is important to consider proper costs to assess the economics instead of technological aspects only. While some scholars are in the opinion that RES only adds complexity — as RES production is volatile and stochastic — others share the opinion that RES has value in terms of energy savings, as well as capacity savings, and contributes to the reliability of an energy system. Instead of analyzing the losses as a result of adding RES, it should be focused on the overall system economics, as this is more important than if more or less electricity is lost for whatever reason. Economy of Scale (EoS) can have significant impact on project evaluation, and therefore should be considered in medium- and long-term planning exercises. As a result of not limiting RES to minimize losses within the system, the most economic systems show penetration for photovoltaic of above 80%, and for Wind of above 55%, on peak demand for both locations. This shows that significant energy out of RES is not used during the winter, as the demand is less than 50% of summer RES, adding value to the system, as RES saves fuel and increases system reliability by decreasing the required electricity from conventional generation. Dependent on the location, between 30 and 40% of peak demand on renewable technologies are required before curtailing takes place. It looks like the concept of Effective Load Carrying Capacity (ELCC) is not that important anymore, as prices for RES came down significantly within the last decades. The RES quality still plays an important role but is getting less important as RES costs are decreasing further. It is expected that also average RES quality site will be more economic in several years.**

## I. Nomenclature

CC .......... Capacity Credit
CSP ......... Concentrated Solar Power
DR .......... Discount Rates
ED .......... Economic Dispatch
ELCC ...... Effective Load Carrying Capability
EoS ......... Economy of Scale
FLH ........ Full Load Hours
GEP ........ Generation Expansion Plan
GT .......... (Open Cycle) Gas Turbine
ISO ......... Independent System Operator
KSA ........ Kingdom of Saudi Arabia
LCoE ...... Levelized Cost of Electricity
LCoF ...... Levelized Cost of Fuel
OPF ........ Optimal Power Flow
O&M ...... Operation and Maintenance
PF ........... Power Flow
PV ........... Photovoltaic
RES ......... Renewable Energy Sources
SCUC ...... Security Constrained Unit Commitment
UC .......... Unit Commitment
VVC ........ Volt/Var control
WACC .... Weighted Average Cost of Capital

## II. Introduction

Capacity credit (CC) of renewable energy sources (RES) is a heavily discussed area of interest within power system planning. There is no industry standard for how to calculate the Effective Load Carrying Capacity (ELCC), which if divided by the added capacity, results in the CC [1] [2] [3]. Some old-established electricity transmission planners, as well as generation expansion planners, are not convinced of the capacity value RES can add to power systems. They argue that RES only adds complexity, as RES production is volatile and stochastic, and therefore cannot be trusted in an energy system where customers expect the system to be available all the time. RES advocates share the opinion that RES has value in terms of fuel savings, as well as capacity displacement, and does contribute to the reliability of an energy system. It is well known that adding renewables above a certain amount, and in the absence of energy storage systems, there is potential for curtailment of RES power plant output at certain moments when demand maybe to low to admit such RES capacity in the system [14]. Therefore after a certain amount of RES is in a system, without or with a limited amount of storage, more and more RES curtailment will be observed. Within some assessed European systems, it was found that the ideal penetration (in terms of RES energy provided) for photovoltaic (PV) is between 8 and 16%, if the distribution network losses are minimized [4]. The challenge of generalization of conclusions with these publications is that every study is system dependent; demand profile, conventional generation technology and reliability targets tend to be unique for each system, specifically in the case of isolated systems, and most of the time only one RES or conventional technology is added for expansion purpose studies, which is not realistic. This means that adding PV, Wind or Concentrated Solar Power (CSP) alone to test the expansion of a system, and assess its economic performance under the same reliability target is not the way to assess this question. It is necessary to add at least some conventional generation at the same time as several RES, to get a better understanding of the implications of adding renewable generators.

It is important to specify the term "renewable penetration," as most of the time it used for the share of electricity demand produced by renewable sources, while it also can refer to the share of renewables capacity compared with either system´s conventional installed capacity or expected peak demand. A penetration of 1% of RES in energy corresponds to different levels of RES capacity penetration, depending on the RES quality, varying from 5% (for load factors arround 0.2, like a good solar PV location) to 2% (for load factors arround 0.5, like excellent wind location).

Another aspect not covered in such exercises found in the literature is that it does not matter how much energy out of the RES power plant is lost in transmission or curtailed, as the system economics for the entire system will naturally answer the question in an appropriate manner. Focusing on distribution and transmission losses are interesting exercises from a technical perspective, but do not add value from a total system economic assessment. An additional aspect within generation expansion planning (GEP) is economy of scale (EoS). EoS is a well-known concept within manufacturing and considers the change in production and maintenance cost when a manufacturer produces more of the same product [5]. Especially in areas far from existing transmission systems RES is already competitive and even within interconnected systems RES is getting more and more attention, as the latest bids within the Middle East show values for energy output of PV systems as low as \$24/MWh or \$0.024/kWh [6].

Within this work, the main question to be answered is "Is it possible to predict the ELCC for a mixed expansion plan, where conventional technologies and renewable energy sources are considered in parallel, while the ELCC calculation is done for PV, Wind, CSP, and GT expansions independently?" This question will be answered while a) assessing the relevance of EoS in the considered GEP framework, b) assessing the importance of RES curtailment and their implications from an economic perspective, and c) if ELCC is an important metric to be considered in GEP (or is it enough that the cost of fuel is higher than the cost of electricity out of RES technologies) going forward, in the light of RES prices getting cheaper.

## III. Literature Review

### A. Use of Effective Load Carrying Capability (ELCC)

Notwithstanding from a definition perspective, Capacity Credit (CC) is the ratio of the system specific ELCC divided by the added name plate capacity of an added technology, the terms CC and ELCC are often used interchangeably [3]. For simplification, both terms are used interchangeably. Regardless of the importance of ELCC in terms of planning generation adequacy, ELCC is not considered in short-term stochastic economic dispatch (ED). In ED forecasted RES generation, historical details are used in stochastic models as inputs, to forecast the most economic operation of available generation resources, while considering the system´s technical limits and possibilities [7]. Such technical limits might include: e.g., transfer limits of transmission lines, up and down ramping capabilities of conventional thermal generation, and minimum up and down time of conventional thermal generation. Other operational models also include optimal power flow (OPF) within their framework, to consider active and reactive power, as well as voltage limits [8] [9]. Important to mention is that not only RES have uncertainty in their production, also the expected load (or demand) has uncertainty, which must be covered within short-term security constrained unit commitment (SCUC). It is also necessary to consider contingencies (N-1 or N-2) or stochastic outages. In short-term (day or weak-ahead) SCUC, ELCC is not included, as RES generation forecast and demand forecast cover the required parameters to optimize the existing system within the planning period [7].

If for medium- and long-term optimization, ELCC is considered that depends on the model purpose. Some models, for e.g., micro grids, only consider some given RES time series and keeps the risk of uncertainty within the RES profile to the utility they are connected to [10]. In general, long-term optimization of ELCC must be considered as RES have the feature to decrease the required conventional thermal power plant expansion, and therefore reduce cost within the utility sector [11]. The challenge within long-term planning is that ELCC is a function of demand profile, RES capacity penetration, quality of available RES, correlation between the hourly profiles of demand and RES, available thermal generation, as well as the overall reliability of a system. Therefore, generic ELCC expectations are not available and ELCC expectations should be estimated for each system individually. Sometimes medium-term models are ignoring active and reactive power, as they try to answer questions such as the optimum storage size, and for such questions this might be acceptable [7]. The trend in medium- and long-term optimization is to consider active and reactive power, as well as the amount of increased RES [12]. This becomes more important with added RES, as also the Volt/Var controlling (VVC) from distribution system operators (DSO) are incorporated more and more into the independent system operator's (ISO) work, to manage the extra high and high voltage transmission system. Within this work, active and reactive power, as well as VVC, is not covered as it has no impact on ELCC, but future work might consider the differences between large scale utility RES, and residential scale systems and their different VVC contribution.

### B. Use of Economy of Scale (EoS)

Most publications until now consider just one or two different sizes and the costs for the considered technologies [10] [13] [14]. Not seldom technology sizes are predefined independently from the load to fulfil. This can result in either too small or too large units, and therefore discriminates some of the defined expansion candidates. In the present work, the search space for the best solutions is widened. Within this work EoS is considered for all covered technologies (PV, Wind, CSP, and GT). The case study in this work will show the importance of allowing EoS within generation expansion planning. In the present article, even EoS for maintenance is considered, as maintenance is assumed as a constant percent factor of invested capital (2% for PV and Wind, 3% for GT and CSP) [15]. Especially for PV and Wind, the EoS discussion is important, as they are a very modular system (e.g., PV panels of 200 W, wind turbines of 2 MW). This modularity is an advantage, as it can be adjusted to needs very easily and precisely.

Without considering EoS, the system´s capacity expansion candidates can be too small or too large, and do not fit into the systems demand requirements. System size from conventional thermal systems also impacts efficiency, while renewable and conventional systems see the impact of maintenance cost depending on size. Previous market potential studies have assumed, for example, customers with peak demand of between 50 kW and 8750 kW, but added conventional technology candidate sizes of 60 to 1000 kW only [10] [13]. By doing so, the candidates do not show an economic fit to the customer demand requirements. This can result in underestimating, as well as overestimating, the economic potentials within the assessed markets (California and the U.S., respectively).

Most long-term investment frameworks work with predefined system sizes [10] [13], or assume a capital cost independent from the installed size of technology, and just assume that large additions will take place (e.g., 500 MW where cost does not change that much anymore) [3] [16]. For countries where significant demand growth is expected, such an assumption might be acceptable, but for mature countries where the demand is not growing or decreasing, such an assumption might result into an underestimation of expected capital requirements. The size of technology to the increased demand or generation to be replaced, is important to be covered to be able to install the most economic addition.

## IV. CASE STUDY

The case study considers two locations: the first one is Sharourah in the Southern Operating Area, and Al Ula in the Western Operating Area. Going forward this case study's locations are referenced as location (1) and location (2).

### A. Assumptions

The costs assumed for the considered technologies (PV, Wind, CSP, and GT) as well as the GT electric efficiency have incorporated EoS effects, and are based on the latest work from Lazard, Lawrence Berkeley National Laboratory, and the Gas Turbine World Handbook, and are shown in the left graph in Figure 2 and Table 1 [15] [17] [18]. The dotted lines show the impact of size on the efficiency to expect for the gas turbine (GT), while the line represents the change of capex ($/Watt). Also, EoS within operation and maintenance (O&M) is considered, as it is assumed that O&M is a constant percent of the invested capital (2% for PV and Wind, 3% for GT and CSP). The square dots at the top of the graph shows the change in capex for CSP, while the dotted line and the dashed dotted lines shows required capex for wind and PV respectively. Figure 2 shows if each technology would have an EoS cost lines, a low cost dotted line, and a high cost dashed line. Especially in the range between $0.5/Watt and $1.5/Watt, it is obvious that selecting the technology cost might have significant impact.

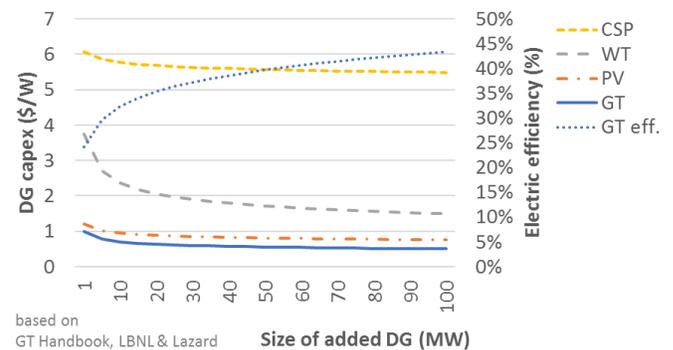

Figure 1: DG cost and efficiency assumptions [15] [18] [30]

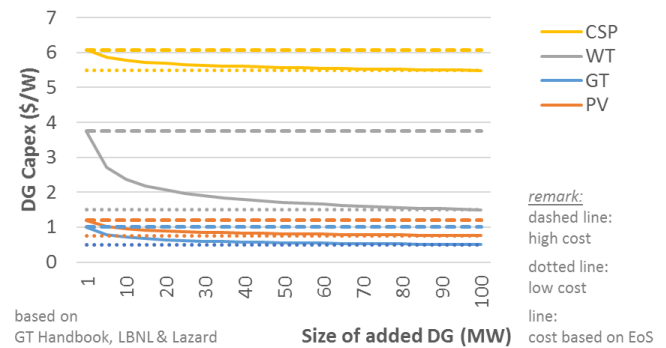

Figure 2: DG capex EoS assumptions [15] [18] [30]

Table 1: Power coefficients for EoS [15] [18] [30]

| power coefficients | PV | WT | GT | CSP | GT HR |
|---|---|---|---|---|---|
| a | 1200 | 3750 | 1000 | 6071 | 14126 |
| b | -0.100 | -0.200 | -0.150 | -0.022 | -0.127 |

The annuity (also known as annuity certain or annual equivalent amount) method used in this work considers a project life time of 20 years, and therefore assumes a depreciation and an expected capital return period of 20 years as well [19]. For projects in industry and the commercial sector, this is usually considered long, while for utility-scale projects, this is still reasonable, and in the case of PV, some could argue 25 years should be the most correct life time. Industry and commercial sector usually invests only if their invested capital can be returned within 3 to 7 years, depending on the industry the company is active on and, as well, other parameters as to how the company is considering different risks in their assessment (risk averse, neutral, or seeking). Therefore, discount rates (DR) of 15-30% (leads to about 3 to 7 years of payback period) would be considered in such cases. The DR should be, in equilibrium, the rate of return that a business could earn if it chooses another investment with equivalent risk. Usually DR coincides with the weighted average cost of capital (WACC), as businesses have access to several sources of funds, as debt and equity with different capital costs.

Moving from capital expenditures (capex) to average electricity generation cost, the importance of EoS will be even more obvious. Figure 3 and Figure 4 shows the cost of electricity considered within this work for new distributed generation technologies with different DR has a significant impact on the economics of a project at the case study location (1). Nevertheless, within project assessments the DR is often used as an indicator of which return a project shall make. While low DR of 6% or lower is favoring investments into renewable energy generation, as these investments are capital intensive (left side), while a higher DR, e.g., 15%, favors conventional technologies as their initial costs are lower than initial renewable cost, and the future costs (especially fuel costs) have less value today. As known, higher fuel costs on the other hand would favor renewable generation, almost independent from the DR selected.

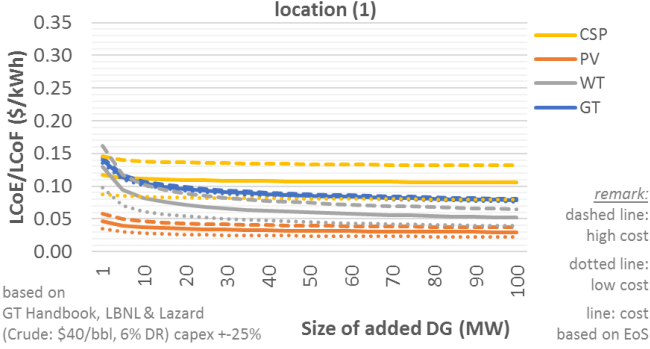

Figure 3: DG electricity generation cost with $40/bbl and 6% discount rate

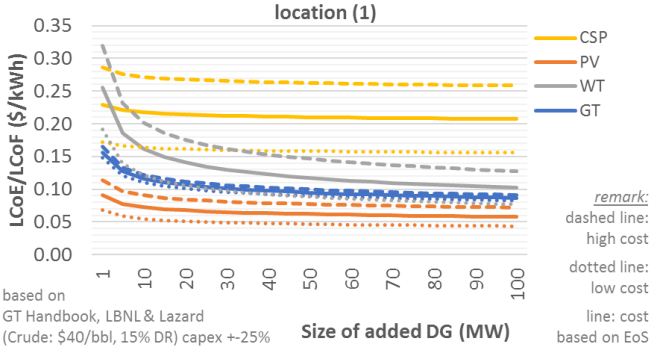

Figure 4: DG electricity generation cost with $40/bbl and 15% discount rate

As shown through Figure 4, fuel value as well as discount rate has a significant role about the question if renewable projects are economic or not. Renewable generation (based on Levelized Cost of Electricity, LCoE) is competitive with existing conventional generation assets, if the displaced fuel (considered as Levelized Cost of Fuel, LCoF) is more expensive [20]. LCoE represents the average electricity generation cost during the entire project life time. Besides the money to spend today (mainly capex), as well as the expected weighted average cost of capital (WACC), other assumptions or estimates like capacity factor, operational and maintenance expenses, and fuel cost and consumption have to be considered as well (see equation (1)).

$$LCoE = \frac{\sum_{n=0}^{N} \frac{C_n}{(1+d)^n}}{\sum_{n=0}^{N} \frac{Q_n}{(1+d)^n}}$$

$$\arg min = \sum_{i=1}^{k} \sum_{x_j \in S_i} \left\| x_j - \mu_i \right\|^2$$

Where $C_n$ is cost from year $n$, $Q_n$ is generated electricity in year $n$, $d$ is the discount rate to be considered, and N is number of years to be considered. For the purpose of this work, a nominal discount rate of 6%, an inflation of 2%, and zero salvage value has been used for the annuity calculations. While LCoE does consider capital expenditures, tax payments, operational and maintenance (O&M), as well as fuel costs, LCoF only considers fuel costs as well as variable O&M costs, to allow a comparison between existing power generation assets and a new built renewable technology asset. With low levels of RES in a power system, it is fair to assume that as soon as the LCoE from new generation is lower than LCoF from existing assets, adding them is beneficial. It is important to mention that both indices (LCoE and LCoF) are simplified concepts, and therefore most of the time do not consider all changes in an existing power system (e.g., ELCC) as well as within power system operation (e.g., changed requirements for spinning and stand-by requirements). Very often the LCoE is used to discuss which project to do and when [21]. But that is not accurate. LCoE represents a levelized cost of the entire project lifetime and assuming predefined production profiles as well as fuel prices. The reality is different: RES resources are varying from year to year, and fuel prices are even more volatile. LCoE is an important indicator to decide which company to go with in the final and last decision, to provide a comparable bidding number for a single technology. LCoE can be used to direct the decision, but should not be used for the final decision. LCoE should not be used to compare different technologies, as they show different types of system improvements (e.g., cost, reliability, fuel savings, full load hours, dispatchability). Neither LCoE nor LCoF considers factors as load following capability, capacity credit, and therefore proper SCUC models covering technical aspects, as e.g., spinning reserve, ramping constraints have to be used to compare different technologies as RES, storage, or conventional generation projects. Therefore, LCoE and LCoF can be used for bid comparison, but are not able to replace a detailed dispatch model assessment.

### B. Methodology

MS-Excel® is used for data handling and easier exchangeability. This work is enhancing a previous work where the initial methodology was enhanced by adding an external economic dispatch (ED) tool into the workflow [22]. Figure 5 shows the updated methodology used within this paper as a control flow diagram. This work added an additional dispatch method where the ED does consider the hourly net load (or residual load) (see circle in Figure 5). This is done by calling the Python based ED tool "Python for Power System Analysis" (pyPSA) [23]. The pyPSA is based on Pyomo, a "Python-based, open-source optimization modeling language with a diverse set of optimization capabilities" [24], and is able to consider static, linear, and security-constrained linear optimal power flow. Pyomo, and therefore pyPSA, can use different commercial and open source linear and nonlinear solvers. For this work the well-known open source solver 'GNU Linear Programming Kit' (GLPK) was used [25].

The variable operational and maintenance (O&M) costs as well as fuel costs are considered within the dispatch, described before where the annuity for new generation (PV, wind, CSP, or GT) as well as replacement costs for existing conventional generation is added to calculate the annual costs. The replacement cost is the cost of replacing an existing generation asset with a new one where today's costs are considered as an annuity. Table 2 shows the proposed technology options within this case study.

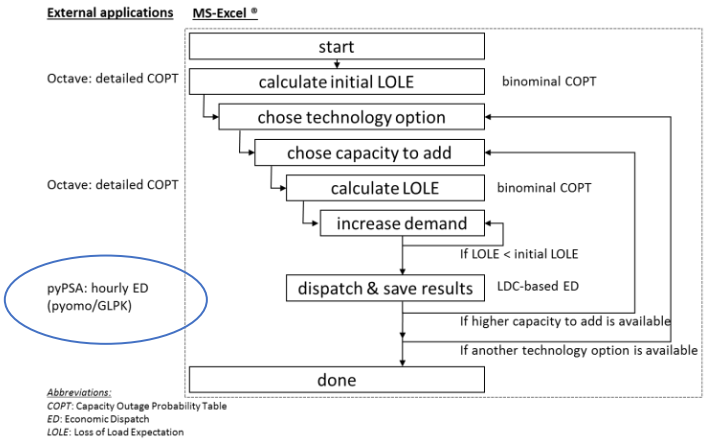

Figure 5: Methodology for reliability analysis and economic dispatch

Considered technology options are pure GT, PV, wind, as well as CSP. Adding renewables at the same time as conventional technology was considered. Therefore between 10 and 50% of the peak demand of conventional technology was added accordingly (see cases 8 to 22). For each of the shown technology options, an addition between 5% and 150% of peak demand (in 5% steps) has been assessed regarding change in reliability within the mini-grid, as well as the expected increase in load to be below the current reliability metric (expressed as LOLE) of the existing system. The 150% limit has been selected to be way above peak demand, to assess the impact of large scale RES without considering storage. It is recognized that RES additions with such high levels might require additional operational instruments, e.g., storage or flexible load by considering demand response and demand side management [26].

### C. Results

Figure 6 to Figure 9 shows a comparison of added capacity and average production cost ELCC and average production cost for locations (1) and (2). The difference is that in location (1), adding Wind does not reduce average production cost as the wind resources there are not good enough. The expected full load hours (FLH) for the renewables at location (1) is for PV 2058, for Wind 2522, and for CSP: 4180, and for location (2) for PV 2199, for Wind 4468, and for CSP: 4179, respectively. Also, the local load profiles are different, so is the peak-to-average ratio for location (1) 1.88 while it is 1.98 for location (2). This is especially interesting as the existing average power generation efficiency is 22% for location (1) and 37% for location (2). Also, available capacity plays a role as location (1) has 125 MW (represents about 120% of peak demand), and location (2) has 75 MW (represents about 150% of peak demand) conventional power generation capacity.

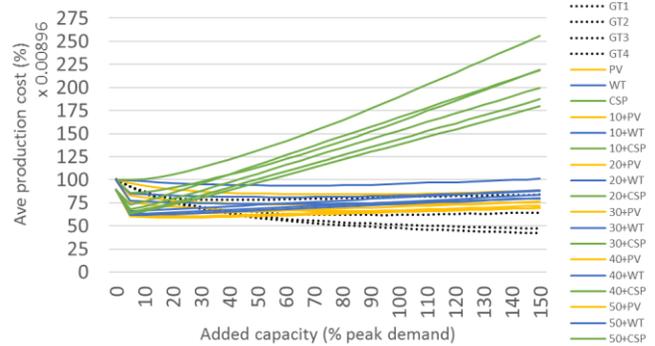

Figure 6: Ave production cost vs. added capacity for location (1)

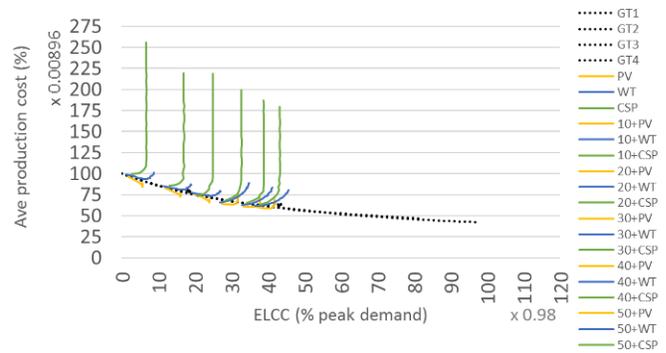

Figure 7: Ave production cost vs. ELCC for location (1)

Figure 10 and Figure 11 shows the same details as above, but zooms into more details. As a result of the lower energy production out of PV, it can be observed that there is a sharp change in average production cost for PV, while Wind has a smoother change in average production cost based on the expected higher FLHs. If location (1) would have RES resources as in location (2) then also Wind would add significant economic value to the overall system. To be more economic than conventional technologies, the alternative with RES has to be below the black dotted lines, which represents adding one to four GTs. Common in both figures is that if the quality of the RES in the according location is good, the average production cost is decreased.

This is not the case by adding Wind in location (1) (see Figure 11). In both locations, PV is able to decrease more than 10% ($9 for location (1) and $5 for location (2), respectively) above the conventional only solution.

Table 2: Considered technology options

| technology option | 1 | 2 | 3 | 4 | 5 | 6 | 7 | 8 | 9 | 10 | 11 | 12 | 13 | 14 | 15 | 16 | 17 | 18 | 19 | 20 | 21 | 22 | 23 | 24 | 25 | 26 | 27 | 28 | 29 | 30 |
|---|---|---|---|---|---|---|---|---|---|---|---|---|---|---|---|---|---|---|---|---|---|---|---|---|---|---|---|---|---|---|
| cost method [1] | size | | | | | | | | | | | | | | | | | | | | | | max | min | max | min | max | min | max | min |
| No. of OCGT's added [2] | 1 | 2 | 3 | 4 | 2 | | | | | | | | | | | | | | | | | | | | | | | | | |
| PV share, (%) | | | | | 100 | | | 100 | | | 100 | | | 100 | | | 100 | | | 100 | | | 100 | | 100 | | | | | |
| WT share, (%) | | | | | | 100 | | | 100 | | | 100 | | | 100 | | | 100 | | | 100 | | | 100 | | 100 | | | | |
| CSP share, (%) | | | | | | | 100 | | | 100 | | | 100 | | | 100 | | | 100 | | | 100 | | | 100 | | 100 | | | |
| OCGT share, (%) | 100 | 100 | 100 | 100 | | | | | | | | | | | | | | | | | | | | | | | | | 100 | 100 |
| OCGT cap. (% peak) [3] | | | | | 10 | 10 | 10 | 20 | 20 | 20 | 30 | 30 | 30 | 40 | 40 | 40 | 50 | 50 | 50 | | | | | | | | | | | |

(1) cost methods are either size, min, or max; size: cost is a function of size (Economy of Scale), min: lowest possible cost, max: highest possible cost considered
(2) capacity of OCGT is splitted into number of units
(3) added OCGT capacity on top of renewable generation; size is a function of peak demand

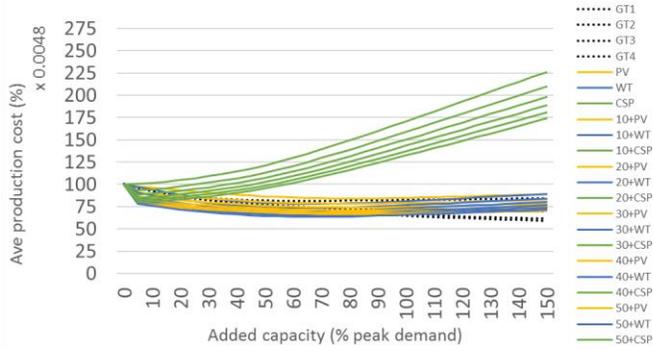

Figure 8: Average production cost for location (2)

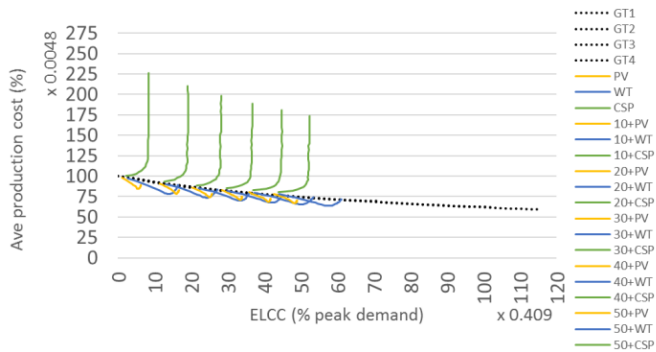

Figure 9: Average production cost for location (2)

In both cases, adding a large amount of PV capacity (150% of peak demand) is more economic than the conventional only case. Similar for Wind, about 120% of peak demand can be added before the average production cost is higher than the conventional only solution.

It is important to mention that the current version of pyPSA does not consider spinning reserve or forecasting errors within the dispatch. Ramp rates have not been considered for this case study, as all generation in both locations are fast reacting GT technologies. The reduction of $4 - $9 allows some increased balancing costs, and still the RES solution would be more economic and is in line with expected balancing costs [31]. As location (1) has less efficient conventional technologies, the resulting price reduction is higher than for location (2) where the available conventional technologies are already efficient. But the RES resources are better in location (2), and therefore the reductions based on RES considerations are higher than for location (1). Nevertheless, as long as the share of RES energy is below 20% for PV and about 40% for Wind, adding RES is decreasing the overall system cost. Also adding GTs above today's situation is improving efficiency, and therefore reducing the system costs. The final decision has to be made case by case, by asking the question: how much peak growth do I see in the short- and medium-term future? The save answer is adding GTs as well as RES, as this results in the cheapest option as long as the RES resources are appropriate. In the case of location (1), only adding PV is beneficial, while for location (2) PV and Wind can be used. So, if available capacity is already satisfying the next year's capacity, adding PV is always the preferred option, as average production costs will come down by doing so. From a risk mitigation perspective, adding RES does displace high cost fuels and GTs can be added fast in a case where demand growth is significant over the subsequent years. In some cases, the minimum RES capacity to be installed is 10% of peak demand, to be more economic than the conventional technology only.

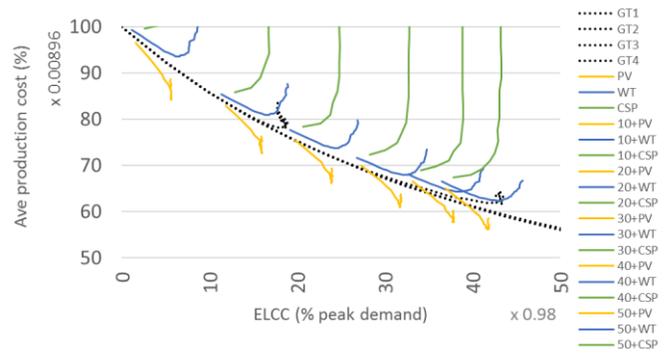

Figure 10: Zoomed in average production cost for location (1) and (2)

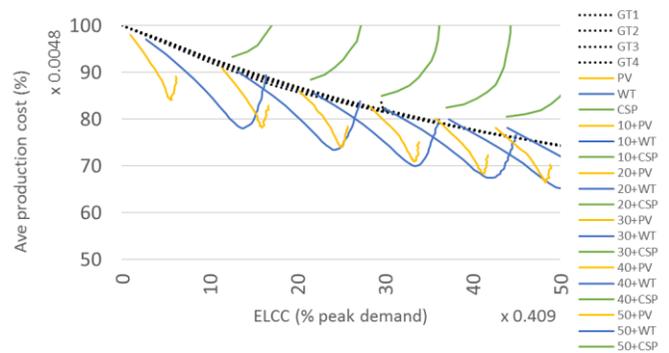

Figure 11: Zoomed in average production cost for location (1) and (2)

Figure 12 and Figure 13 shows the impact on average production cost and RES production share within location (1) and (2). The most economic PV penetration for location (1) is between 85 and 90% of the peak penetration, while for location (2) it is between 80 and 85% of peak penetration. Wind has an optimal penetration level of between 55 and 65% of peak demand for location (1), while location (2) shows an optimal penetration level of 65% for all scenarios. Using data from location (2) in location (1) would result in economic Wind scenarios, and would add 70 to 75% of peak demand as most economical. Dependent on the location, it needs between 30 and 40% of peak demand on renewable technologies before curtailing is considered at all.

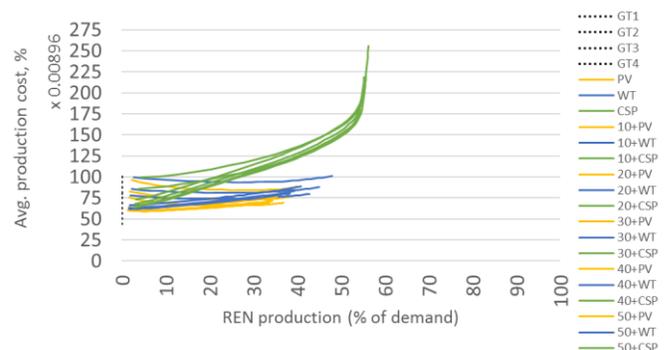

Figure 12: Average production cost vs. share of RES production

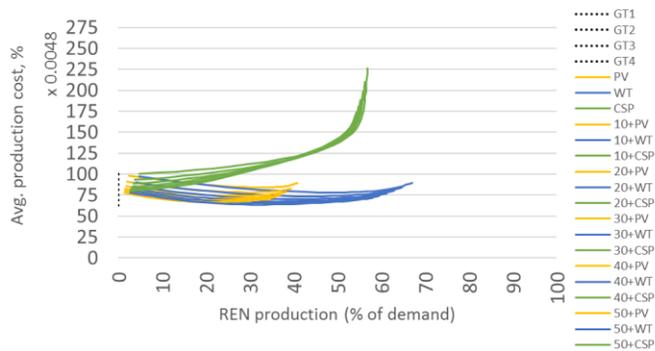

Figure 13: Average production cost vs. share of RES production

## V. Conclusions

While CC and ELCC is a heavily discussed topic within power system planning and operation, it would help to have a standardized way of considering such details in short- and long-term planning exercises. While some scholars are of the opinion that RES only adds complexity, as RES production is volatile and stochastic; others share the opinion that RES has value in terms of energy savings, as well as capacity savings, and does contribute to the reliability of an energy system. Instead of analyzing the losses as result of adding RES, the focus should be on the overall system economics, as this is more important than if more or less electricity is lost. EoS can have significant impacts on project evaluation and therefore should be considered in medium- and long-term planning exercises. As a result of not limiting RES to minimize losses within the system, the most economic systems show penetration for PV of above 80%, and for Wind of above 55% on peak demand for both locations. This shows that even significant energy out of RES is not used during winter, as the demand is less than 50% of summer RES, add value to the system, as they save fuel and increase the system reliability by decreasing the required electricity from conventional generation. Dependent on the location, it needs between 30 and 40% of peak demand on renewable technologies before curtailing is considered at all. It looks like the concept of ELCC is not that important anymore, as prices for RES came down significantly within recent years. From a fuel savings perspective, they are getting more important and economical. The RES quality still plays an important role but is getting less important as RES costs are decreasing further. The current work does not consider logistic costs for fuel. Considering them may result in WT being economic even in location (1).

It is important to mention that this work is based on a single year of data only, not considering ramp constraints and spinning reserve, and therefore is valid from a directional point of view, but might see differences if a multiyear assessment was undertaken.

## VI. Future Work

Future work might shed more light on GEP by also adding storage on top of RES, to analyse if that could further decrease the average electricity cost for the given locations. Next steps also could be to incorporate EoS as well as detailed spinning requirements (as a result of forecasting errors for demand and RES) in an open source GEP software package, e.g.,

OSeMOSYS or pyPSA [27]. Adding RES capacity above peak demand is critical, and needs more detailed work to understand the detailed implications and consequences. The current version of pyPSA is not able to consider spinning reserve or forecasting errors within the dispatch procedure which might be an area of improvement going forward. An alternative would be to increase the variable costs for RES, to incorporate some balancing costs once the installed RES capacity is significant [31].

## VII. Acknowledgments

The authors gratefully acknowledge valuable comments and support from colleagues and management within Saudi Aramco (especially from Power Systems Renewables and Power Systems Planning).

## IX. Biographies

**Markus Groissböck** has been a Power Generation Expansion Planner for about 10 years. His research interests lie in medium- and long-term investment planning within the power and energy sector, considering distributed and variable energy resource planning and energy demand forecasting.
(email: markus.groissboeck@aramco.com)

**Alexandre Gusmao** has been a Renewable Energy Business Developer for more than 15 years, covering a portfolio of projects from wind, PV, hydro and biomass. His interests are in planning for renewable integration in power grid and mini-grids, forecasting natural energy resources, forecasting assets prices.
(email: alexandre.gusmao@aramco.com)

APPENDIX: SUPPLEMENTARY MATERIALS

Figure 14 shows the normalized average daily profiles for location (1) and (2), respectively. The base data show a peak demand for location (1) of 84 MW and for location (2) of 35 MW.

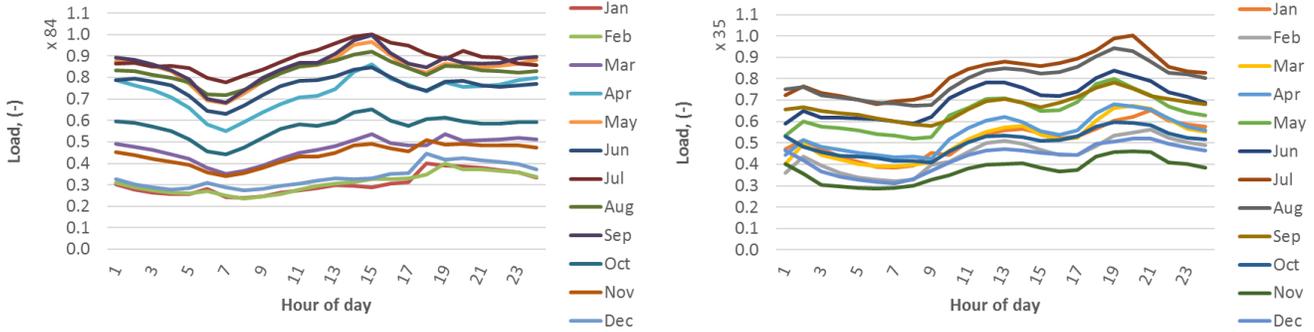

Figure 14: Normalized daily average demand for location (1) and location (2)

Within carbon dioxide emission reduction, a common call for a DR of 0% is required to speed up a transition to a sustainable energy system [28]. Figure 15 shows what that would mean for a $15/bbl and $30/bbl scenario, while considering a DR of 0%. With the $15/bbl scenario, large PV and Wind are economic compared with a new built conventional GT plant. The $30/bbl scenario would see all renewable technologies as economical. This shows that with a 0% DR $30/bbl are enough to favor RES over conventional technologies. Therefore, it is important to make a wise and reasonable decision about which value to consider for DR. In practice, the decision about the DR is made based on cost of debt, as well as the country risk, where policy risk is the major contributor [29]. DR for onshore wind projects can be as low as 3.5% (Germany), 5.7% (France) as well as 10% or more (e.g., Hungary, Romania, and Croatia) can be found in Europe, as the country risk drivers, as well as the financing possibilities.

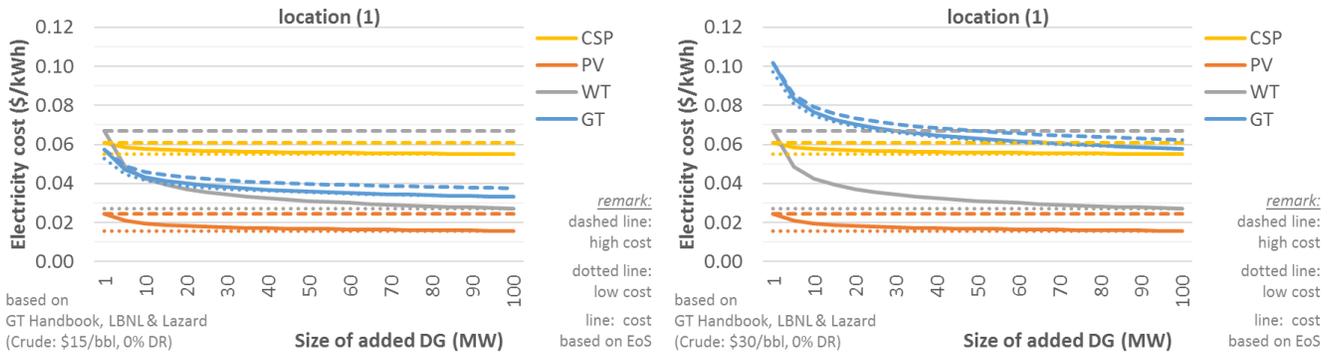

Figure 15: DG electricity generation cost with 0% DR and 15$/bbl or 30$/bbl

Each dotted blue line in Figure 16 represents different fuel price assumptions, and range between $10 and $100 per barrel. In the case that future payments are not discounted (0% DR) $30/bbl are enough to let the LCoF be more expensive than renewables LCoE. If future payments are discounted with 6% (which represents something like a low end for utility scale projects) more than $60/bbl is required to be more expensive than all renewable technologies, while $30/bbl is required for Wind and PV to be more economic than existing conventional technologies.

Figure 17 shows how LCoE and LCoF looks like with renewable resources as available in location (2). While in location (1) (see Figure 16) PV would be preferred all the time. In location (2) there is almost no difference from a levelized cost perspective. While location (1) needs $30/bbl within a 6% DR assessment location, location (2) is more economic even in a $10/bbl environment.

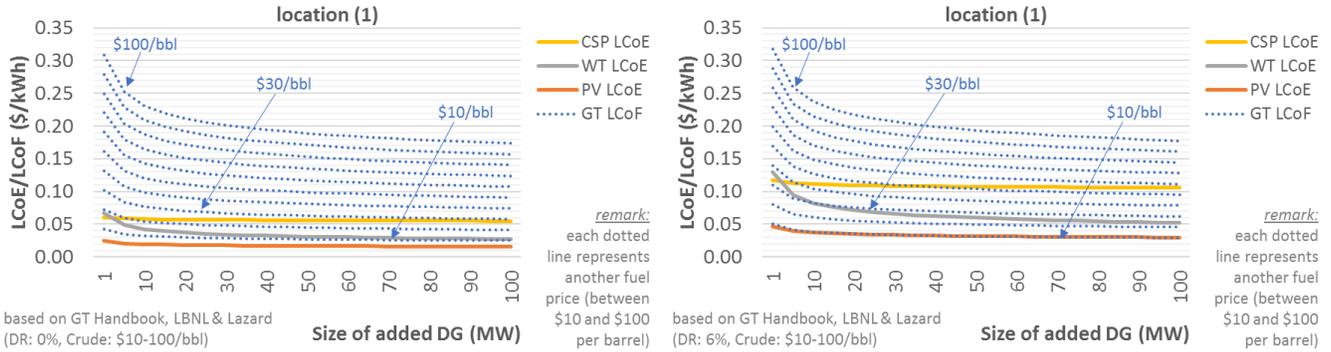

Figure 16: LCoE vs LCoF with DR of 0% and 6% for location (1)

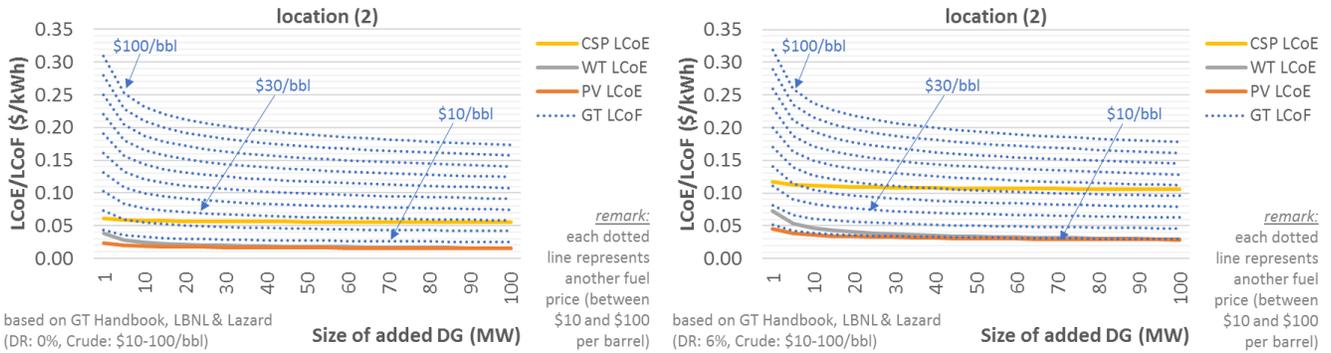

Figure 17: LCoE vs LCoF with DR of 0% and 6% for location (2)

While it is crucial to consider fuel price impacts on the economics, it is as crucial to consider other uncertainties or risks as well (e.g., financing, resource, environmental, regulatory, offtake, and technology uncertainty) [30]. Capex assumptions of ±25% are considered as a sensitivity in Figure 18.

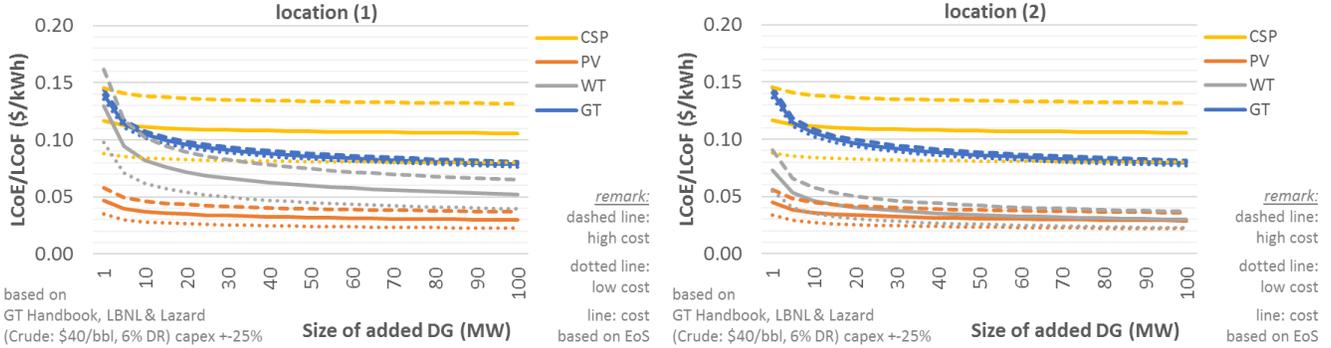

Figure 18: Capex sensitivity (+- 25%) for location (1) and (2)

Figure 19 and Figure 20 shows a comparison of ELCC and CC as well as Annuity and CC for location (1) and (2). For comparison reasons, all figures within the result section are normalized, based on peak demand, annuity without any additional investments, and average production cost, respectively. Figure 19 (for location (1)) shows very similar behavior for both locations and all technologies (GT, PV, Wind, and CSP). Figure 20 (for location (2)) shows a similar behavior for GTs and CPSs, but not for PV and Wind. In this comparison, all technology additions are compared with a factious 100% reliable generation technology. In reality, no technology is 100% reliable, and therefore the differential between an assessed conventional and an assessed renewable technology has to be made.

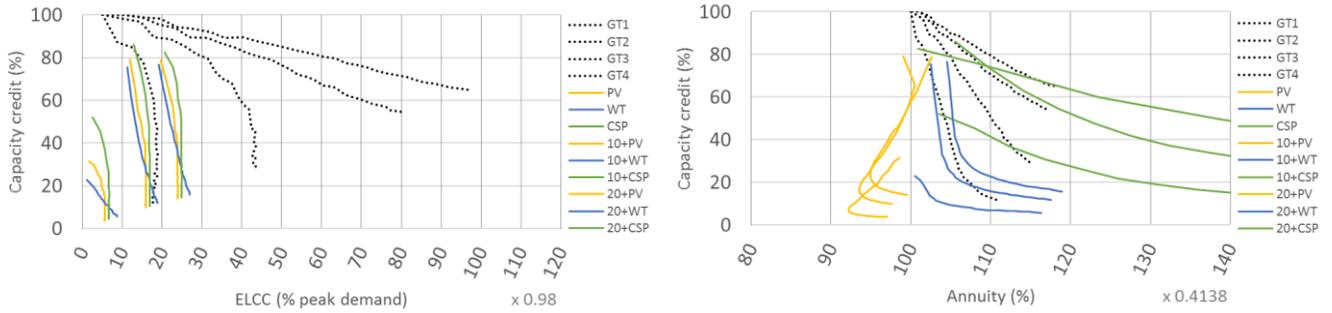

Figure 19: Capacity credit and costs for location (1)

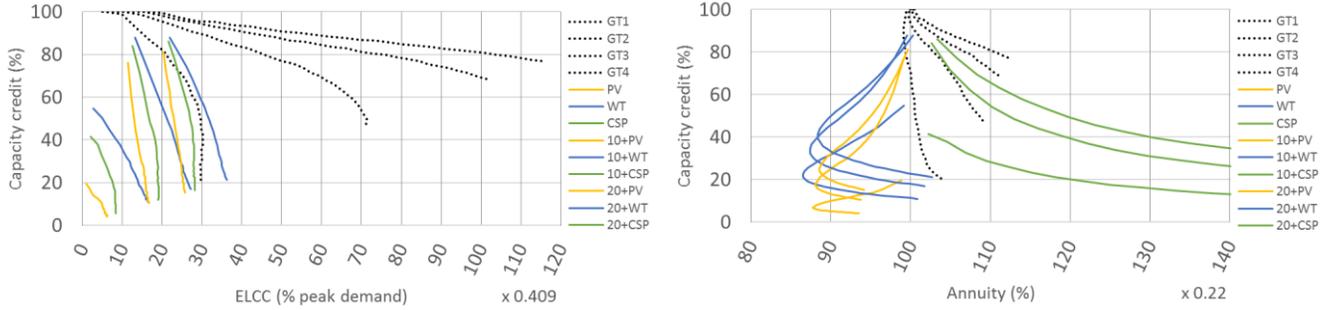

Figure 20: Capacity credit and costs for location (2)